\documentclass[useAMS,usenatbib,usegraphicx]{mn2e}
\title[Injection energy of positrons]{Restrictions on the injection energy of positrons annihilating near the Galactic center}
\author[Chernyshov et al.]{D.O.~Chernyshov$^{1,2,3,4}$, K.-S.~Cheng$^2$, V.A.~Dogiel$^{1,4}$, C.-M.~Ko$^{4}$, and W.-H. Ip$^{4}$\\
$^1$I.E.Tamm Theoretical Physics Division of P.N.Lebedev Institute, Leninskii pr, 53, 119991 Moscow, Russia\\
$^2$Department of Physics, University of Hong Kong, Pokfulam Road, Hong Kong, China\\
$^3$Moscow Institute of Physics and Technology, Institutskii lane,
141700
Moscow Region, Dolgoprudnii, Russia\\
$^4$Institute of Astronomy, National Central University, Jhongli
320, Taiwan\\}

\begin{document}
\label{firstpage}

\maketitle

\begin{abstract}
The origin and properties of the source of positrons annihilating in the Galactic Center is still a mystery. One of the criterion, which may discriminate between different mechanisms of positron production there, is the positron energy injection. Beacom and Y\"uksel (2006) suggested a method to estimate this energy from the ratio of the 511 keV line to the MeV in-flight annihilation fluxes. From the COMPTEL data they derived that the maximum injection energy of positron should be about several MeV that cut down significantly a class of models of positron origin in the GC assuming that positrons lose their energy by Coulomb collisions only. However, observations show that the strength of magnetic field in the GC is much higher than in other parts of the Galaxy, and it may range there from 100 $\mu$G to several mG. In these conditions, synchrotron losses of positrons are significant that extends the range of acceptable values of positron injection energy. We show that if positrons injection in the GC is non-stationary and magnetic field is higher than 0.4 mG both radio and gamma-ray restrictions permit their energy to be higher than several GeV.
\end{abstract}


\section{Introduction}
One of the interesting and still unsolved problems is the origin
of 511 keV annihilation emission from the Galactic Bulge. It is
observed as an extended diffuse emission from $5^\circ - 8^\circ$
radius region  with the flux $\sim 8\times 10^{-4}$ ph
cm$^{-2}$s$^{-1}$ that requires the rate of positron production
there
 $\sim 10^{43}$ s$^{-1}$ \citep{knoed,chur,jean}.
These observations showed that the energy of annihilating
positrons was about 1 eV. On the other hand, all potential sources
of positrons in the Galaxy like SN stars \citep{knoed}, massive
stars generating the radioactive $^{26}$Al \citep{prantzos},
secondary positrons from p-p collisions \citep{cheng2}, lepton
jets of AGNs \citep{tot}, dark matter annihilation
\citep{boehm,sizun}, microquasars and X-ray binaries
\citep{weid,silk} etc. generate positrons with energies $\ga 1$
MeV. This means that positrons should effectively lose their
energy before annihilation and thus generate emission in other
than 511 keV energy ranges. Therefore, the injection energy of
positrons is an essential parameter for modeling annihilation
processes, and it can be in principle discriminated from
observations.

High energy positrons annihilate "in-flight", thus producing
continuum emission in the range $E>511$ keV. A prominent 511 keV
line emission is generated by these positrons when their energy is
decreased to the thermal one due to energy losses. Therefore, one
can expect that the continuum and line emission are proportional
to each other.

 For the  lifetime
of in-flight annihilation, $\tau_{if}$, and the average cooling time
$\tau_{cl}$ of high energy positrons, one can estimate the
expected flux of in-flight annihilation, $F_{\mbox{1-30 MeV}}$
from the observed 511 keV line emission, $I_{line}$, as
\begin{equation}\label{tt_by}
F_{1-30~\mbox{MeV}} \sim
\frac{\tau_{cl}}{\tau_{if}}I_{line}~~~~\mbox{if $\tau_{if}\gg
\tau_{cl}$}
\end{equation}
where
\begin{equation}
\tau_{if}=\left[n\sigma_{if}(E)v(E)\right]^{-1}
\end{equation}
and
\begin{equation}
\tau_{cl}=\int\limits_{E_{inj}}^{E_{th}}\frac{dE}{\left(dE/dt\right)_{cl}}
\end{equation}
Here $n$ is the plasma density, and  $E_{inj}$,
$E_{th}$ and $v(E)$ are the
injection energy of positrons, the energy of thermal plasma, and
the positron velocity, respectively, $\sigma_{if}$ is the
cross-section of  in-flight annihilation. The function
$(dE/dt)_{cl}$ is the rate of energy losses defined as sum of
Coulomb, synchrotron, inverse Compton, bremsstrahlung etc. losses:
\begin{eqnarray}
&&\left(dE/dt\right)_{cl}=\left(dE/dt\right)_{coul}+
\left(dE/dt\right)_{syn}+\\
&&+\left(dE/dt\right)_{IC}+\left(dE/dt\right)_{br}+...\nonumber
\end{eqnarray}

If cooling of the positrons is only due to the Coulomb losses,
then $\tau_{if}$ and $\tau_{cl}$ are proportional to $n^{-1}$, and the
relation (\ref{tt_by}) is independent of the medium density. So,
this ratio of the continuum in-flight and the annihilation
emission is universal and can be applied even to a medium with an
unknown density. \citet{by2006} assumed that positrons in the
Galactic center (GC) loose their energy by Coulomb interactions
only, and they suggested to use this ratio for the analysis of the
annihilating positron origin in the GC. In the above-mentioned
models the injection energy of positrons is expected in the range
from several  to hundreds MeV. Therefore, the in-flight gamma-ray
emission is also expected in this energy range.

The MeV flux from the central part of the Galaxy  was
observed by COMPTEL \citep[see,][]{compt}.  The origin of this
emission is still unclear since the known processes of gamma-ray
production (like inverse Compton, bremsstrahlung etc.) are unable
to generate the observed flux \citep{strong, portt}.
\citet{cheng2} assumed that this excess in the GC direction might
be due to the in-flight annihilation of fast positrons. However,
it is observed not only in the direction of the GC. The excess is
almost constant along the Galactic disk \citep{compt} where the
intensity of annihilation emission is lower than in the Galactic
centre. This makes problematic the in-flight interpretation of
this excess in the disk since the ratio 511 keV flux/in-flight
continuum is constant. If the in-flight flux is responsible for
the MeV excess in a relatively narrow central region ($\la
5^\circ$), absolutely the same excess in other parts of the disk
remains unexplained \citep[see][]{sizun,chern1}.

Therefore, in \citet{by2006} and latter in \citet{sizun}  a more
firm constraint on the in-flight gamma-ray flux from the Galactic
center was suggested. According to their criterion the in-flight
flux should not exceed several statistical errors of the COMPTEL
measurements. That gives an upper limit for the injection energy
about several MeV. Then models assuming higher injection energy
should undoubtedly be rejected.

Below we show that under some conditions the injection energy may be
higher than 10 MeV in contrast to conclusions made in papers mentioned
above, and, thus, there is a room for models assuming
injection of high energy positrons.

Thus, \citet{cheng1,cheng2} assumed that these positrons are
secondary and generated by collisions of relativistic protons
injected from black hole jets. The theoretical analysis of
\cite{ist} confirmed the hadronic origin of jets and showed that
protons were accelerated there by the stochastic and the
centrifugal acceleration up to energies $E_p\simeq 10^{20}$ eV
that might offer an explanation to the recent results of the
Pierre Auger collaboration \citep{auger}. If such or similar
mechanism produces indeed enough relativistic protons with Lorentz
factor   $\gamma \ga 2$ in the vicinity of the central black hole
then we do expect there an effective production of secondary
positrons with energies above 30 MeV, just as assumed in
\cite{cheng1,cheng2}.

Processes of $p-p$ collisions produce also a flux of
gamma-rays in the range above 100 MeV by decay of
$\pi^\circ$-meson, and below this energy by, so-called, internal
bremsstrahlung radiation of secondary electrons \citep[see for
detail][]{haya}.  A flux of gamma-rays in the 1 to 30 MeV range
from internal bremsstrahlung may be  higher than the mentioned
in-flight flux. Thus, \citet{bbb} showed that the internal
bremsstrahlung flux is  very significant, if positrons
 in the GC are generated by dark matter annihilation. However,
 in the dark matter model positron production in the GC
is stationary. On the other hand, from the restrictions derived
from  EGRET data it follows that the positron production in the GC
should be strongly non-stationary, if these positrons are
generated by  $p-p$ collisions \citep{cheng1, cheng2}. The flux of
gamma-rays from $p-p$ collisions is significant during a very
short period after a star accretion onto the black hole. At
present this flux has decreased in several orders of magnitude
from its initial value and, therefore, is unseen.

 Below in section \ref{sc_mag}
we shall show that the condition of non-stationarity is also
required to fit radio observations.

\section{Medium properties in the vicinity of the Galactic center}
As follows from observations, the central  200 pc region of the
Galaxy is strongly nonuniform. The inner bulge (200-300 pc)
contains $(7-9)\times 10^7~M_\odot$ of hydrogen gas. In spite of
relatively small radius this region contains about 10\% of the
Galaxy's molecular mass. Most of the molecular gas  is contained
in very compact clouds of mass $10^4-10^6M_\odot$, average
densities $\geq 10^4$cm$^{-3}$.

However, this molecular gas occupies a rather small part of the
central region, most of   which is filled with a very hot gas.
ASCA \cite{koya1} measured the X-ray spectrum in the inner 150 pc
region which exhibited a number of emission lines from highly
ionized elements which are characteristics for a $8-10$ keV
 plasma with the density 0.4 cm$^{-3}$. Later on Chandra
observations of \cite{muno} showed an intensive X-ray emission at the
energy $E_x\sim 8$ keV from the inner 20 pc of the Galaxy. The
plasma density was estimated in limits $0.1-0.2$ cm$^{-3}$. Recent
SUZAKU measurements of the 6.9/6.7 keV iron line ratio
\citep{koya2} was naturally explained by a thermal emission of 6.5
keV-temperature plasma.

One should note that there is no consensus on the magnetic field
strength in the GC. Estimations ranges from about or smaller than
hundred $\mu$G \citep[see][]{spergel, radio, higdon}, up to several
mG (\citealt{plante, yuza1}, see in this respect the review of
\citealt{ferri}). Radio observations of the central regions show that
the structure of strong magnetic fields is nonuniform  and it
concentrates in filaments which extends up to 200 pc from the GC.
The region containing mG magnetic field is estimated by the
angular size $1.5\degr \times 0.5\degr$ \citep[see
e.g.][]{morris1, yuza2, morris}.

\begin{figure}
\center
\includegraphics[width=0.5\textwidth]{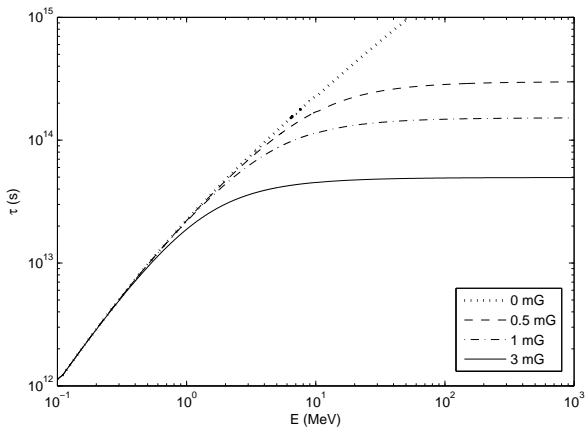}
\caption{Characteristic cooling times of positrons for different values of magnetic field strength.}
\label{lifetm}
\end{figure}

 In these magnetic fields
synchrotron losses are essential  even for positrons with energies
3-30 MeV. One can see from Fig. \ref{lifetm} that if the magnetic
field strength is as high as  3 mG, the cooling time of
 1 GeV positrons is the same as that of positrons with
energies about 1 MeV experiencing only Coulomb losses.

Structure of the magnetic field and the way electrons and positrons
interact with it is also unknown. In our model we assumed that
positrons interact with magnetic field structures violently thus
allowing us to average the magnetic field over central 100 pc radius.
However if the interaction is weak and can be neglected the original
constraints deduced by \citet{by2006} prevail.

We shall show that the injection energy of positrons can be much
higher than 1 MeV  because of synchrotron losses in the Galactic
center that extends significantly the class of models  explaining
the origin of  annihilation emission from the bulge.

If it is not specified  we accepted below that the
central 100 pc radius region is filled with strong magnetic field
 and the gas density in the central region
($r\la 500$ pc) is $\ga 0.2$ cm$^{-3}$. The rate of
ionization losses depends on the medium ionization degree.
Therefore, we consider two cases of neutral and fully ionized
medium.

\section{Effects of Strong Magnetic Fields in the GC}\label{sc_mag}

If positrons propagate through the region of strong magnetic
fields in the GC, they effectively lose their energy and emit
radiation in the radio range of frequencies. Thus, for $H = 3$ mG
positrons with energies 10-100 MeV generate emission in the range

\begin{equation}
\nu_0 = 0.29\frac{3eH}{4\pi mc} \left(\frac{E}{mc^2}\right)^2 \sim
1-100~\mbox{MHz}
\end{equation}

For the magnetic field of 3 mG the rate of synchrotron losses of
relativistic electrons and positrons is higher than for any other
process of losses. Therefore, an intensive synchrotron emission is
expected from this region.

Observations of the GC gave the following values of radio flux
from there \citep[see, e.g.,][]{mezger,radio}. This flux is about
10 kJy in the frequency range 300 - 700 MHz for the central region
$3^\circ\times 2^\circ$. For the region $1.5^\circ\times
0.5^\circ$ this flux is almost one order of the magnitude smaller
at the frequency 330 MHz, i.e. $\sim 1$ kJy.

On the other hand, simple estimations of synchrotron emission
shows that if electrons and positrons are injected with energies
about 100 MeV, then the radio flux from the GC is much higher than
1 kJy.

Below we demonstrate this point from very simple estimates of the
radio emission from the GC. In the stationary state the total
spectrum of positron in the GC injected at the energy $E_0$ with
the rate $Q$ is
\begin{equation}
N(E)\simeq \frac{Q}{|dE/dt|}\theta(E_0-E)
\end{equation}
where the rate of synchrotron losses is
\begin{equation}
\frac{dE}{dt}=-\frac{2}{3}\frac{e^4H^2}{c^3m^2}\left(\frac{E}{mc^2}\right)^2=-\beta
E^2 \label{loss}
\end{equation}
Here and below $N(E)$ is the density of electrons and positrons with energy E.
The radio flux from the GC at Earth can be calculated from the
integral
\begin{equation}
\Phi_\nu=\frac{1}{4\pi d^2}\int\limits_E N(E)P(\nu,E)dE
\end{equation}
where $P(\nu,E)$ is the emissivity of a single positron with the
energy $E$, and $d=8$ kpc is the distance between Earth and the
GC. To get an estimation we use the approximation for the function
$P$ from \citet{ber90}
\begin{equation}
P(\nu,E)\simeq \frac{dE}{dt}\delta\left(\nu -\nu_0(E)\right)
\end{equation}
Then we have the the radio flux at the frequency $\nu$
\begin{equation}
\Phi_\nu=\frac{Q}{8\pi
d^2}\frac{mc^2}{\nu}\sqrt{{\nu}/{\bar{\nu}}}
\end{equation}
where $\bar{\nu}=\sqrt{0.29\cdot 3eH/4\pi mc}$. For the production
rate of secondary electrons $Q=2\cdot 10^{43}$s$^{-1}$, this
equation gives a flux at Earth about $10^9$ Jy at the
frequency $\nu\sim 100$ MHz that is several order of
magnitude higher than observed. This means that either the
magnetic field strength and the injection energy of positrons
should be smaller than those used in this estimation or the
situation is strongly non-stationary.

For estimations of the non-stationary case we use  the source
function in the form
\begin{equation}
Q(E,t)=N_0\delta(t)\delta(E-E_0)
\label{delta_inj}
\end{equation}
where $N_0$ is the total number of injected positrons. Then the
distribution function of positrons is (see section \ref{evol})
\begin{equation}
N(E,t)=N_0\delta\left(E-\frac{E_0}{\beta E_0t+1}\right)
\end{equation}
 where $\beta$ is the coefficient in Eq. (\ref{loss}), and the
 radio flux has the form
\begin{equation}
\Phi_\nu\sim\frac{N_0}{\nu{\left(E=\frac{E_0}{\beta
E_0t+1}\right)}}\left.\frac{dE}{dt}\right|_{E=\frac{E_0}{\beta
E_0t+1}} \label{nu_sh}
\end{equation}
As one can see that the current energy of electrons is independent
of their injection energy $E_0$ and equals $E\simeq 1/\beta t$ for
long enough $t$.

From Eq.(\ref{nu_sh})  it follows that the radio flux just after
the injection of positrons equals $\sim 3 \cdot 10^7$ Jy for the
$N_0\simeq 2\cdot 10^{55}$ in each capture events that is
necessary to produce the observed annihilation flux from the GC.
  However,  for the time $t\simeq 10^5$ years (the average period
  between str captures) the peak of intensity
 shifts from the frequency 100 MHz to several MHz where this radio flux
 cannot be observed because of absorption in the interstellar gas,
 and just this effect is a key point of our analysis presented
 below.

 As an example we show in Fig. \ref{s_radio} the spectrum
radio emission  from the GC region produced by secondary electrons
and positrons  at the time $10^5$ yr after the capture. The
magnetic field strength in the GC is 2 and 3 mG. As one can see
for long time after the capture the peak of emission shifts from
the frequency of hundreds MHz to several MHz, and the intensity of
radio emission is negligible at 330 MHz. For calculations of radio
emission we used the accurate equation
 \citep[from e.g.][]{ber90},
\begin{equation}
\Phi(\nu,t)_{s} = \frac{1}{4\pi d^2}\frac{\sqrt{3}e^3H}{mc^2} \int
\limits_0^\infty N(E,t)dE\frac{\nu}{\nu_c} \int
\limits_{\nu/\nu_c}^\infty dz K_{5/3} (z) \label{radio_int}
\end{equation}
where $N(E)$ is the total number of electrons and positrons with the energy $E$,
$d=8$ kpc is the distance to the GC, and $K_\alpha(x)$ is the
McDonald function.

\begin{figure} \center
\includegraphics[width=0.5\textwidth]{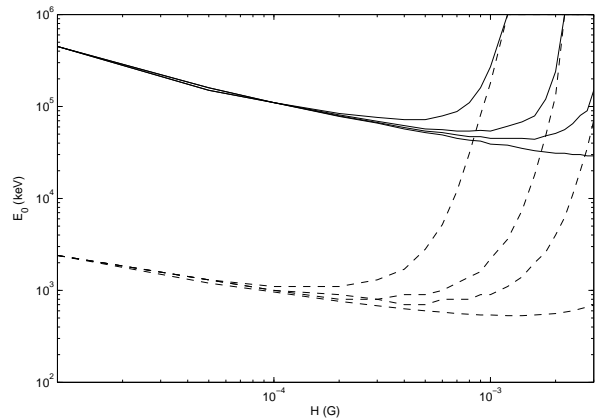}
\caption{ Limitations from radio observations for different values of time T passed since capture.
In ascending order T = $1\times 10^4$, $5\times 10^4$, $10^5$, and $2\times 10^5$ years.
Solid lines correspond to injection in form of delta-function, dashed lines correspond to power-law
injection.} \label{lim_radio}
\end{figure}

In Fig. \ref{lim_radio} we presented the limitations of
positron injection energy $E_0$ derived from radio data  for
different values of the magnetic field strength and different
 times T passed from the last star capture. We  considered two
possible cases: the injection in the form of delta-function,
 Eq. (\ref{delta_inj}), shown in the figure by solid
lines and  power-law injection spectrum (dashed line):
\begin{equation}
N_0(E)=N_0\delta(t)E^{-3}\theta(E-E_0) ~\mbox{.}\label{sp_radio}
\end{equation}
One can see that if the injection is non-stationary, then the
injection energy of positrons can be extremely high   in case of
strong magnetic fields in the GC. Indeed for almost stationary
situation when characteristic period between injections is small
$T \leq 1\times10^4$ yr the maximum allowed energy in case of
$H \geq 1$ mG cannot exceed 30 MeV due to radio limitations. In case
of power-law spectrum situation is even worse. However if period
is long enough and magnetic field is strong the situation changes:
for $T=1\times 10^5$ yr and $H=2$ mG the maximum energy is about 1 GeV
and it can be even higher for longer periods and higher values of
the magnetic field.

\begin{figure} \center
\includegraphics[width=0.5\textwidth]{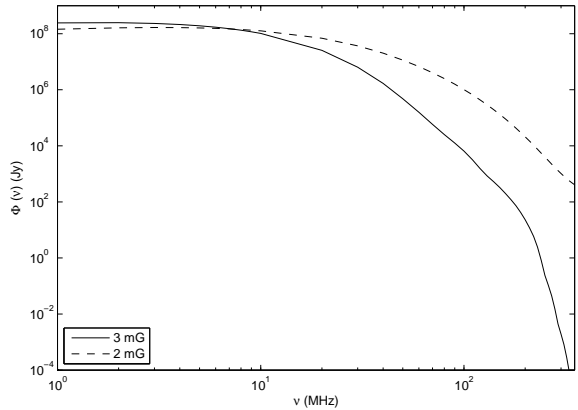}
\caption{ Radio flux at Earth produced by positrons from a single
capture event  at the time $10^5$ yr after the capture. The
strength of magnetic field $H=3$ mG (solid line) and $H=2$ mG
(dashed line)} \label{s_radio}
\end{figure}

\section{Constraints of the Model of Annihilation Emission}
There are restrictions for our model which follow from radio
and gamma-observations:
\begin{enumerate}
\item The flux of radio emission from the GC at present should not exceed the value of 1 kJy in
the range 300 - 700 MHz \citep[see, e.g.,][]{mezger} that
restricts the number of high energy positrons in the GC. The
radius of high strength magnetic field sphere was taken to be
$r_H=100$ pc;
\item The flux of annihilation emission observed at
Earth from this region is $8\cdot 10^{-4}$ photons
cm$^{-2}$s$^{-1}$ for the FWHM=$6\degr$ central region
\citep{chur};
\item The flux in the MeV energy range was measured by COMPTEL.
 As in \citet{by2006,sizun} we accept that the flux of
in-flight annihilation  emission should not exceed the COMPTEL
$2\sigma$ level of  in the energy range 1-30 MeV, if it is due to
in-flight annihilation.
\end{enumerate}

These  restrictions are presented in  Table \ref{param_st1}.
\begin{table}
 \caption{Constraints of the model}
 \centering
\begin{tabular}{|p{2.9cm}p{2.6cm}p{1.8cm}|}
\hline
$\Phi_{\nu}(\nu=330$ MHz)&$F_{1-30~\mbox{MeV}}$&$I_{511~\mbox{keV}}$\\
$1.5\degr\times 0.5\degr$ region&& FWHM =$6\degr$\\
 \hline
$1$&$<2\sigma$ of COMPTEL&$8\cdot 10^{-4}$ \\
kJy&&ph
cm$^{-2}$s$^{-1}$\\
 \hline\label{param_st1}
\end{tabular}
\end{table}

\section{Evolution of the Spectrum of Relativistic Positrons}\label{evol}
We start from the spatially uniform model, which estimates the
total flux of gamma-rays and radio emission (i.e. integrated over
the volume of emission). The evolution of positron spectrum  can
be described by the equation
\begin{equation}
\frac{\partial N}{\partial t}+\frac{\partial }{\partial E}
\left(\frac{dE}{dt}N\right)= Q(E,t)\label{N_density}
\end{equation}
where $dE/dt$ is the rate of synchrotron losses (see Eq.
(\ref{loss})). In case of instantaneous injection $Q(E,t) =
N_0(E)\delta(t)$ where $N_0(E)$ is the injection spectrum of
positron,  the solution of this equation is
\begin{equation}\label{inp_spectrum1}
N=\frac{1}{|dE/dt|}N_0\left(\tau-t\right)
\end{equation}
where
\begin{equation}
\tau=\int\limits_E^\infty\frac{dE}{|dE/dt|}
\end{equation}
and $dE/dt$ is the rate of synchrotron or Coulomb losses.

In the high energy range, where the synchrotron losses are
essential, the spectrum is
\begin{equation}\label{inp_spectrum2}
N = \frac{1}{(1-\beta E t)^2}N_0\left({\frac{E}{1-\beta E t}}\right)\theta(1-\beta E t)
\end{equation}
We see that independent of the injection spectrum the maximum
energy of  positrons long after the injection  is
\begin{equation}
E_{max}\simeq\frac{1}{\beta t}
 \label{max}
\end{equation}
 where  $t$ is time after the injection.

\begin{figure}
\center
\includegraphics[width=0.5\textwidth]{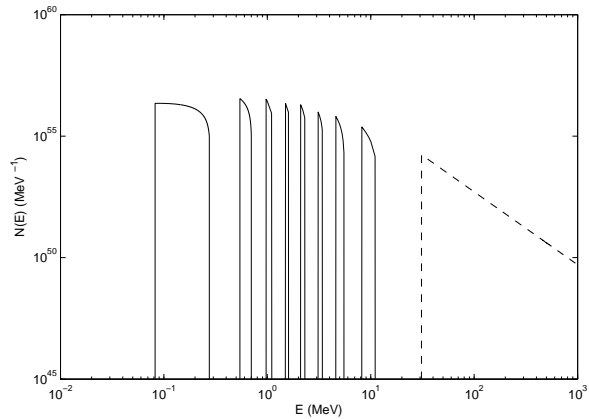}
\caption{ Spectrum of secondary positrons  as a result of several
successive capture events. The strength of magnetic field $H=3$
mG.} \label{bunch}
\end{figure}

If the positrons are secondary, then their injection spectrum has
a cut-off in the low energy range at $E_{\rm cut} \simeq 30$ MeV
because of the threshold of $p-p$ reaction, and the energy
distribution of positrons leaving the region of strong magnetic
fields looks like a succession of separated bunches (see
Fig.\ref{bunch}). Here and below we take the injection spectrum of
secondary positrons  in the form
\begin{equation}
N_0(E)=A_bE^{-3}\theta(E-30~\mbox{MeV}) \label{spectrum}
\end{equation}
where $A_b\simeq 5.4\cdot 10^{58}$ MeV$^2$ gives the total
number of injected positrons $\sim 3\cdot 10^{55}$, where the
spectral index  equals $-3$ as  in the Galactic disk
\citep[see][]{ber90}.

The bunch width is
\begin{equation}\label{bunch_width}
\Delta E\sim \frac{1}{(\beta t)^2E_{\rm cut}}~\mbox{.}
\end{equation}
The time duration of the bunch (characteristic time during which
 the bunch crosses any energy  $E$ under the influence of
 synchrotron losses) is independent of  $E$
 and for the spectrum (\ref{spectrum}) equals
 \begin{equation}
 T_b=\frac{1}{\beta E_{\rm cut}}
 \end{equation}
 that gives $T_b=3.6\cdot 10^4$yr for $E_{cut}=30$ MeV and $H=3$
 mG. Subsequent bunches are separated from each other by the periods of
 $T_i\sim 10^5$ yr,  the average periods of star capture.

 After the capture time equaled $10^5$ yr the maximum energy of
 positrons in the bunch
 shifts from $\infty$ to $E\simeq 10$ MeV (see Eq. (\ref{max}), and there are no positrons in the energy range
 $E> 10$ MeV while the energy range below 10 MeV contains permanently
 a number of bunches moving to the thermal region (see Fig. \ref{bunch}). Then we
 conclude that the spectrum of positrons in the range above 10 MeV
 is strongly non-stationary (shown by the dashed line in Fig. \ref{bunch}), as well as the emission
 generated by these particles. In contrary, the spectrum of positrons with energies
$\la 10$ MeV (shown by the solid lines in Fig. \ref{bunch}) and
the radiation produced by these positrons is quasi-stationary.

 When the rate of synchrotron losses drops down with positron energy,
the following evolution of positrons in the range $E<1$ MeV occurs
under the influence of Coulomb losses. The rate of Coulomb losses
is \citep{haya, ginz}
\begin{equation}
{{dE}\over{dt}}=-{{2\pi
e^4n}\over{mc\beta(E)}}\ln \Lambda \label{iot}
\end{equation}
where $\log \Lambda$ is Coulomb logarithm, and
$\beta(E)=v/c$. For lorenz-factor $\gamma = \frac{E}{mc^2}+1$ in a neutral medium
\begin{equation}\label{log_neu}
\log \Lambda \sim \log \left[ (\gamma-1)(\gamma^2-1)\right] + 20.5 ~\mbox{,}
\end{equation}
while in completely ionized plasma
\begin{equation}\label{log_ion}
\log \Lambda \sim \log \left[ \gamma/n\right] + 73.6 ~\mbox{.}
\end{equation}

\section{Emission Produced by Non-Thermal Positrons}

Relativistic positrons with energies higher than 10 MeV generate
radio emission due to synchrotron losses and fluxes of in-flight
annihilation and bremsstrahlung emission in the energy range above
10 MeV.

\begin{figure}
\center
\includegraphics[width=0.5\textwidth]{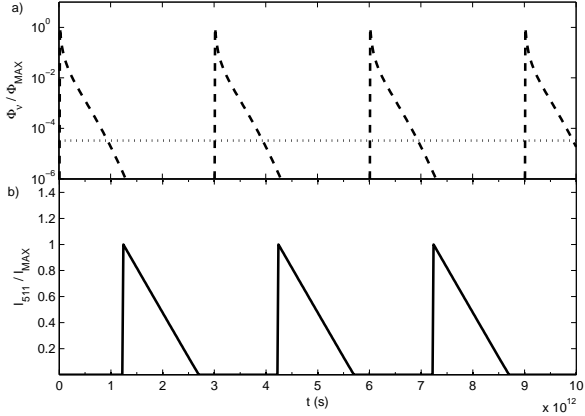}
\caption{a) Time-variations of the radio flux from the GC. Here
$\Phi_{max}^{radio}=3\cdot 10^7$ Jy. b) Annihilation flux
variations from the GC in the framework of non-stationary model.
Here $I_{max}^{511}=8\cdot 10^{-4}$ photons cm$^{-2}$s$^{-1}$.}
\label{n-st}
\end{figure}

From Eqs. (\ref{radio_int}), (\ref{N_density}) and
(\ref{spectrum})
 we calculated the time variations of radio
emission from the GC at the frequency 330 MHz. The result of these
calculations is shown in Fig. \ref{n-st}a. As it was expected at
the very beginning, just after the injection of high energy
positrons, the flux of radio emission is as high as
$\Phi_{max}^{radio}\sim 3\cdot 10^7$ Jy, but its value decreases
very rapidly with time. The present time corresponds to periods
when the radio flux drops down below the level about 1 kJy (dotted
line), as the observations require.

\begin{figure}
\center
\includegraphics[width=0.5\textwidth]{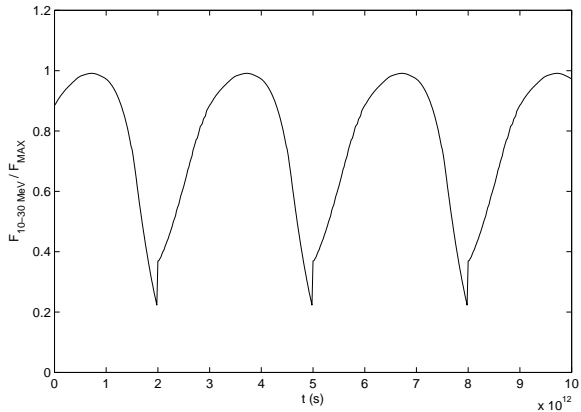}
\caption{ Variations of the in-flight 10-30 MeV flux. The value of
$F_{max}$ is presented in Table \ref{outp_st}.}
\label{comptel_var}
\end{figure}

The in-flight annihilation in the range 10-30 MeV is  produced by
positrons whose intensity is non-stationary. Therefore, this
flux from the GC is time-variable (see Fig. \ref{comptel_var})

On the contrary, the flux of in-flight annihilation in the ranges
1-3 and 3-10 MeV is almost stationary as expected.

The  in-flight fluxes calculated  for the cases of neutral and
ionized medium and the COMPTEL $2\sigma$ levels are presented in
Table \ref{outp_st}. The flux of the in-flight annihilation in the
range 10-30 MeV presented in the table corresponds to the moment
when the radio at 330 MHz drops to
 1 kJy.

\begin{table}
 \caption{Flux of in-flight annihilation in case of non-stationary annihilation (photons cm$^{-2}$s$^{-1}$).}
 \centering
\begin{tabular}{|p{1.5cm}p{2.0cm}p{1.5cm}p{1.5cm}|}
\hline
Energy range       (MeV)&COMPTEL $2\sigma$&Ionized&Neutral\\
\hline

1-3  & $2\times 10^{-4}$ & $1.2\times 10^{-5}$  & $2.6\times
10^{-5}$
\\
3-10  & $9\times 10^{-5}$ & $3.2\times 10^{-6}$  & $3.6\times
10^{-6}$
\\
10-30  & $3\times 10^{-5}$ & $3.7\times 10^{-7}$   & $3.6\times
10^{-7}$
\\
 \hline\label{outp_st}
\end{tabular}
\end{table}
The bremsstrahlung flux in the energy range from 1 to 30 MeV was
calculated for the cross-section from \citet{haug}. Its values
ranges from $5\times 10^{-7}$ in the range $1-3$ Mev to $3\times
10^{-7}$ at $3-10$ MeV and $1\times 10^{-7}$ at $10-30$ MeV.
The flux units are the same as in Table \ref{outp_st}.
One can see that bremsstrahlung doesn't contribute much since
lorentz-factor is rather small.

\begin{table}
 \caption{Flux of in-flight annihilation in case of stationary annihilation (photons cm$^{-2}$s$^{-1}$).}
 \centering
\begin{tabular}{|p{1.5cm}p{2.0cm}p{1.5cm}p{1.5cm}|}
\hline
Energy range       (MeV)&COMPTEL $2\sigma$&Ionized&Neutral\\
\hline

1-3  & $2\times 10^{-4}$ & $4.6\times 10^{-5}$  & $1\times
10^{-4}$
\\
3-10  & $9\times 10^{-5}$ & $1.2\times 10^{-5}$  & $1.4\times
10^{-5}$
\\
10-30  & $3\times 10^{-5}$ & $1.5\times 10^{-6}$   & $1.4\times
10^{-6}$
\\
 \hline\label{outp_ststat}
\end{tabular}
\end{table}

\begin{figure}
\center
\includegraphics[width=0.5\textwidth]{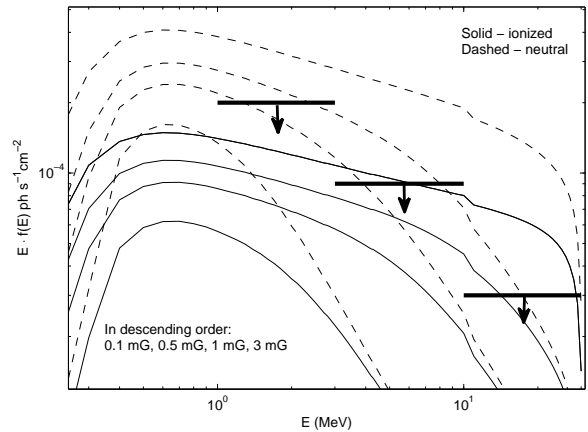}
\caption{Combined in-flight annihilation and bremsstrahlung spectra for
different values of magnetic field. COMPTEL limits are shown by heavy
solid lines.} \label{if_spectrum}
\end{figure}

From these tables one can see that in all energy ranges the
calculated in-flight flux is below the COMPTEL $2\sigma$ level.
Therefore, in contrast to the conclusion of \citet{by2006} and
\citet{sizun}, we find that there is no problem with the injection
energy of positron in the model with strong magnetic field at GC.
Corresponding combined spectra of in-flight annihilation and bremsstrahlung
emission for magnetic fields in the range $0.1-3$ mG are shown in
Fig. \ref{if_spectrum}.

\section{Annihilation Emission of Thermal Positrons}
The characteristic time of  annihilation for thermal positrons
can be defined from
\begin{equation}
T_{ann}=\left[2\int\limits_{E_T}^{\infty}dE\sqrt{\frac{E}{\pi
\Theta^3}}\exp\left(-\frac{E}{\Theta}\right)n \sigma (E)v\right]^{-1}
\label{t_ann}
\end{equation}
 which is a function of
the probability for thermal positrons to be in the energy range
where the annihilation cross-section is high. Here temperature
$\Theta \sim 1$ eV as it usually assumed.
 For the gas density $n\ga 1$ cm$^{-3}$
the annihilation time is about $\la 2.6\cdot 10^4$ yr
\citep[see][]{guessoum} which is smaller than
the average time of star capture $T_i$ and the bunch time $T_b$.
Then the process of annihilation emission  is non-stationary since
all positrons from a bunch annihilate during the time shorter than
the period between two neighbour capture events.

The nondimensional equation for positrons has the form
(here and below $f(p)$ is the density of positrons with momentum $p$)
\begin{eqnarray}
&&\frac{\partial f}{\partial t}+\frac{\partial }{\partial
p}\left(\left.\frac{dp}{dt}\right|_Cf-D_{C}\frac{\partial f}{\partial
p}\right)+\nonumber\\
&&+n_0v\sigma_{if}f+n_0v\sigma_{ce}f=Q(p,t)
 \label{eq_c}
\end{eqnarray}
where $\left.dp/dt\right|_C$  and $D_C$ are the rate of Coulomb
losses and the momentum diffusion due to Coulomb collisions. These
two terms form the Maxwellian spectrum of thermal particles. The
cross-sections $\sigma_{if}$ and $\sigma_{ce}$ denote the
in-flight annihilation of fast positrons and the charge-exchange
annihilation process of thermal particles.

The  fluctuations of the annihilation flux expected in this case
are shown in Fig. \ref{n-st}b. The maximum intensity of the
annihilation emission corresponds to the value of
$I_{max}^{511}=8\cdot 10^{-4}$ photons cm$^{-2}$s$^{-1}$. From
this figure one can see that there are long periods of time when
the annihilation flux reaches its maximum value while the radio
emission is below the 1 kJy level as the observations require. It
means that according to this model we see the annihilation flux
from the bulge at its peak value.

If the gas density  $n_0\la 0.25$ cm$^{-3}$ then the annihilation
time $T_{ann}$ is larger than $T_i$ and the situation is
quasi-stationary as follows from Eq. (\ref{t_ann}). Since the
positrons from a single bunch do not  annihilate completely for
the period between two neighbour injections, then these positrons
 are accumulated in the thermal pool up to
the level which provides a quasi-stationary rate of annihilation.

\begin{figure}
\center
\includegraphics[width=0.5\textwidth]{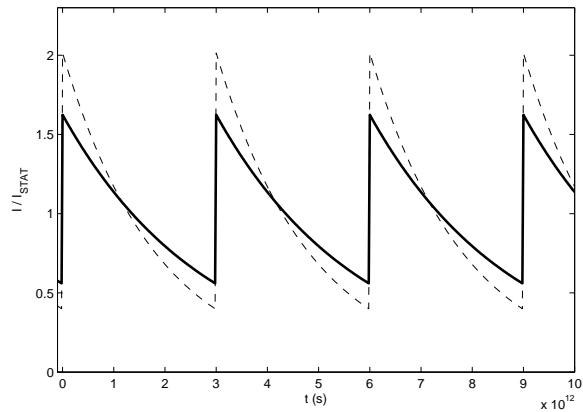}
\caption{ Variations of the annihilation flux from the GC in the
framework of quasi-stationary model: solid line - without the
effect of streaming instability, dashed line  - with the effect of
streaming instability. $I_{stat}$ is the average
value of the annihilation flux.} \label{ann_var}
\end{figure}

From Eq. (\ref{eq_c}) we calculated the flux of annihilation in
the quasi-stationary case which is shown in Fig. \ref{ann_var} by
the solid line. We see that  fluctuations of the annihilation flux
is much smaller than in the non-stationary case (see Fig. \ref{n-st}b).

The expected quasi stationary fluxes  are presented in Table
\ref{outp_ststat}. The bremsstrahlung flux in this case is:
$2\times10^{-6}$ at 1-3 MeV, $1\times10^{-6}$ at 3-10 MeV, and
$3\times10^{-7}$ at 10-30 MeV. Again fluxes are below COMPTEL
$2\sigma$.

\begin{figure}
\center
\includegraphics[width=0.5\textwidth]{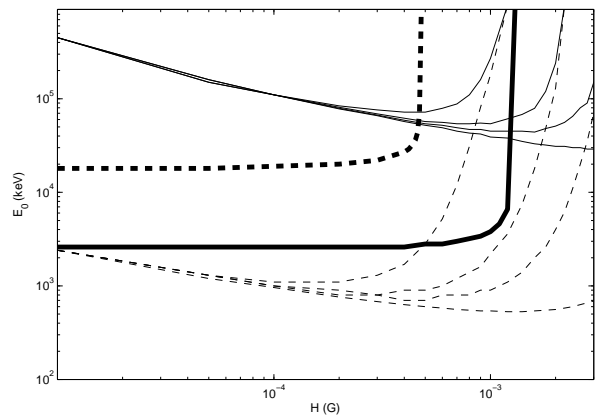}
\caption{The multi-band limitations on the injection energy of positrons.
Thin lines are the same as in Fig. \ref{lim_radio}. Thick dashed line represents
limitations from COMPTEL energy range in case ionized medium. Thick solid lines
represents limitations from COMPTEL energy range in case of neutral medium.} \label{all_limits}
\end{figure}

An important conclusion from our calculations  is that
the only restriction of the model is the rate of synchrotron
losses which should cool down positrons up to the energy $\la 10$
MeV for the time $\la 10^5$ yr, see Fig. \ref{lifetm}. The
regions of permitted values of positron injection energies derived
from the COMPTEL restriction for different values of the magnetic
field strength are shown in Fig. \ref{all_limits}: the thick
dashed line - ionized medium and the thick solid line - neutral
medium. These region is below these lines. As one can see there
are no restrictions for the injection energy  from gamma-ray data
if the magnetic field strength is high enough.

However, as we discussed already in the case of high magnetic
field we have an additional restriction - the flux of radio
emission at 330 MHz.  The range of energy values which are derived
from the radio data are shown by thin lines. In that part Fig.
\ref{all_limits} is the same as Fig. \ref{lim_radio}: each line correspond
to different value of period of injections (in ascending order T =
$1\times 10^4$, $5\times 10^4$, $10^5$, and $2\times 10^5$ years).
Thin solid lines correspond to injection in form of delta-function while
thin dashed line correspond to power-law injection.
Again as we mentioned already for strong magnetic field if period
between two captures is long enough the injection energy can be very high.

So combining the gamma and radio data we conclude that  the
injection energy of positrons can be high  if the model satisfies
the tow conditions: the magnetic field strength is larger than 0.4
mG and the average period between two neighbour captures is longer
than $10^5$ yr.

\section{Landau Damping: Quasi-Stationary Model}
\begin{figure}
\center
\includegraphics[width=0.5\textwidth]{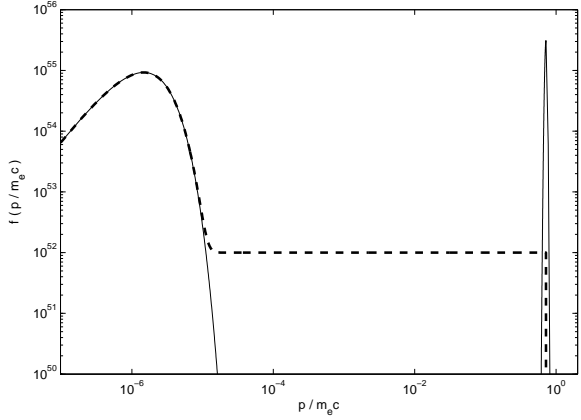}
\caption{ Transformation of the bunch spectrum by the streaming
instability.} \label{stream}
\end{figure}

As it follows from Eq. (\ref{inp_spectrum2}) when a bunch of MeV
positrons escapes the strong magnetic field
region into the surrounding medium the positron distribution
 looks like a two-peak function with one maximum at
thermal energies and the other at the energies of the bunch (see
Fig. \ref{stream}). From the quasi-linear theory it follows that
such a particle distribution is unstable (\citealp[Ch.I, \S 1.16,
p. 104]{artsim}, see also \citealp[Ch.III, \S30, p.243]{lifsh}),
and the flux of fast particles excites effectively plasma waves
due the  streaming instability .
As a result a bunch of fast particles with the energy
$E=E_{bunch}$ is transformed into "a plateau" distribution in the
energy region $E<E_{bunch}$. This process of the bunch smearing
occurs due to resonant wave excitation ($v=\omega_k/k$, where $v$
if the particle velocity and $k$ is the wave number). This process
is described as diffusion in the velocity space with the diffusion
coefficient $D_{st}$ equaled
\begin{equation}
D_{st}\sim
4\pi^2\frac{e^2}{m^2}n_bm\frac{v^2}{\omega_p}\frac{v-v_{min}}{v_{max}}
\end{equation}
where $n_b$ and $v_{max}$ is the density and the velocity of bunch
particles, $v_{min}$ is about the velocity of thermal particles
and $v$ is the current velocity of particles. The Langmuir
frequency $\omega_p$ is
\begin{equation}
\omega_p=\sqrt{4\pi n_0e^2/m}
\end{equation}
where $n_0$ is the density of background plasma. As numerical
calculations showed (see, e.g. \citealp[Ch.I, \S 1.16, p.
107]{artsim}) the bunch smearing was reached for several Langmuir
times that in comparison with the model characteristic times  is
almost instantly. Calculations show that about one third of the
bunch energy is transformed into the energy of excited plasma
waves.

The nondimensional equation for  nonthermal positrons has the
form  in this case
\begin{eqnarray}
&&\frac{\partial f}{\partial t}+\frac{\partial }{\partial
p}\left(\left.\frac{\partial p}{\partial
t}\right|_Cf-(D_C+D_{st})\frac{\partial f}{\partial
p}\right)+\nonumber\\
&&+n_0v\sigma_{if}f+n_0v\sigma_{ce}f=Q(p,t)
\end{eqnarray}

 The expected  variations of annihilation flux from the GC are
shown in Fig. \ref{ann_var} by the dashed line. On can see that
effect of streaming instability does not change the flux value
significantly.

\section{Spatial non-uniform model}

We presented above our analysis of integrated fluxes from the GC.
As we noticed mG  magnetic fields occupy the inner sphere with the
radius $r_H=100$ pc (i.e. its angular radius is $<1\degr$) while
the FWHM of the annihilation emission is about $6\degr$. It means
that in most of their lifetime positrons spend outside the central
magnetic sphere. On the other hand, it follows from our analysis
that positrons should be cooled down by  strong magnetic fields up
to the energy $\la 10$ MeV. These circumstances give restrictions
for processes of positron propagation in the GC.

Spatial variation of the positron distribution function
 $N$ with  propagation terms is described
 by the equation
\citep[see][]{ber90}:
\begin{equation}
\label{fp_main}
- \frac{1}{r^2}\frac{\partial}{dr}r^2D_{rr} \frac{\partial
N}{\partial r} + \frac{\partial}{\partial E}\left[
\frac{dE}{dt}N\right] + n\sigma vN = Q(E,r) ~\mbox{,}
\end{equation}
Here ${dE}/{dt}$  is the rate of total energy losses,  and
$D_{rr}$ is a spatial diffusion coefficient. We use a simple
approximation for the spatial part of the injection function
$Q(r)$, assuming that positron sources are uniformly distributed
inside the sphere of the radius $r_0$, $Q(r,E)=Q(E)\theta(r_0-r)$.
 The function $Q(E)$ is
given by Eq. (\ref{spectrum}) but with the coefficient $A_b\simeq
5\cdot 10^{46}$ MeV$^2$/s as a result of averaging of injection
processes over time (quasi-stationary injection).

The boundary condition are
\begin{equation}
\left(\nabla_r \cdot N\right)_{r=0}=0,~~~~~~~~~N_{r=\infty}=0
\end{equation}
and
\begin{equation}
\left(\nabla_p \cdot N\right)_{E=0}=0
\end{equation}

The loss term can be presented as
\begin{eqnarray}
&&\frac{dE}{dt} = \left(dE/dt\right)_{coul} + \left(dE/dt\right)_{IC} +
\nonumber \\
&& + \left(dE/dt\right)_{synBK} \times \theta (r - r_H) + \nonumber \\
&& + \left(dE/dt\right)_{synS}\times \theta (r_H - r)~\mbox{,}
\end{eqnarray}
where $\left(dE/dt\right)_{synBK}$ is the rate of synchrotron
losses in the outer sphere $r>r_H$ where the magnetic field
strength equals  $H \sim 10^{-5}$ G, and
$\left(dE/dt\right)_{synS}$ is the rate of energy losses inside
the central sphere $r<r_H$ losses in strong magnetic field $H = 3$
mG.

The observed $6\degr$ FWHM of spatial distribution of annihilation
emission corresponds to a sphere with the radius about 400 pc that
requires the spatial diffusion coefficient be about $D_{rr}\sim
10^{28}$cm$^2$s$^{-1}$. We notice here that propagation of MeV
positrons in the GC is still very questionable \citep[see
e.g.][]{jean09}.

\begin{figure}
\center
\includegraphics[width=0.5\textwidth]{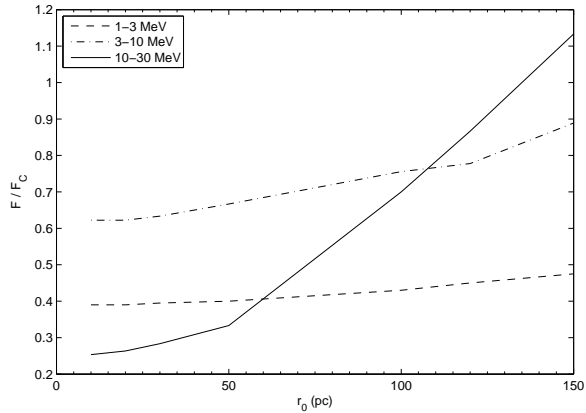}
\caption{ Variations of the in-flight  flux as a function of the
source region radius. Here $F_c$ is COMPTEL's $2\sigma$ level.} \label{rdep}
\end{figure}

Fast cooling by synchrotron losses is possible only inside the
sphere of strong magnetic field $r < r_H$. Therefore, the radius
of source region should be quite small, $r_0 < r_H$.
 Only in
this case positrons are cooled to MeV energies by synchrotron
losses and then fill the 400-pc sphere. In Fig. \ref{rdep} we
presented the calculated values of in-flight fluxes in the three
COMPTEL energy ranges as a function of the source radius $r_0$.
These fluxes  are normalized to the COMPTEL 2$\sigma$ level
presented in Table \ref{outp_st}. As one can see extended sources
with $r_0
 r_H$ are unable to satisfy the COMPTEL restrictions if the
injection energy is 30 MeV.

\section{Conclusion}
We showed that:
\begin{itemize}
\item Unlike
\citet{by2006} and \citet{sizun}, who restricted the injection
energy of annihilating positrons at the value $1-7$ MeV, we show
that under the condition of strong magnetic fields $H \ga 0.4$ mG
in the Galactic center this value can be much larger. This extends
significantly a class of models acceptable for explanation of the
annihilation emission from the GC. However, the necessary
condition in this case is that the positrons should be cooled down
by the synchrotron losses up to the energy $\la 10$ MeV. Otherwise
the expected radio flux and in-flight MeV emission from the GC
exceed the level required from observations;
\item The main restriction of this model follows from the value of observed  radio flux from the
GC which equals 1 kJy at the frequency 330 MHz. In order to
satisfy this condition the peak of annihilation emission should be
shifted in time to the moment when the radio flux falls down below
the observed level;
\item The energy spectrum of annihilating positrons looks like a
number of bunches produced by subsequent capture events. In the
case of the secondary origin of positrons the structure of the
bunch spectrum is almost quasi-stationary in the energy range $\la
10$ MeV, and strongly non-stationary in the range above 10 MeV;
\item On the other hand, the expected flux of the annihilation emission is
quasi-stationary if the density of the background gas is $\la
0.25$ cm$^{-3}$.
\item Time variations of the in-flight annihilation in the energy
ranges 1-3, 3-10, and 10-30 MeV are quite small, and the important
point is, their values are smaller than the COMPTEL
 $2\sigma$ level;
 \item
 Our calculations show that the size of the source region should
 be smaller than the radius of the sphere filled with strong
 magnetic field. Otherwise the in-flight flux exceeds the COMPTEL
 $2\sigma$ level.
 \end{itemize}
It would be interesting to find traces of similar processes
from other galaxies. If positrons injection is related to stellar
capture events as it was suggested by \cite{cheng1} and
\cite{cheng2} one can expect to find some traces of these
processes  in galactic nuclei with recent capture events in the
form of high radio flux. However, galactic nuclei with high X-ray
flux that indicates active accretion processes there do not
show such a powerful radio emission  \citep{wong}. This  is not
surprising since the peak of radio emission is expected in our
model some time after the capture. The delay time is about $10^3$
years as follows from the characteristic time of $p-p$ collisions
which produce high energy electrons:
\begin{equation}
t_{pp} = (n\sigma_{pp}c)^{-1} \sim 10^3~\mbox{yr}
\end{equation}
 for the gas density of molecular clouds  $n\sim 10^4$ cm$^{-3}$.

The flux of X-ray emission from the accretion disk scales with time like
\citep{rees}
\begin{equation}
L_{x} \sim L_{edd}\left(\frac{t}{t_{min}}\right)^{-5/3}
\end{equation}
where $t_{min} \simeq 0.2$ yr. So by the time radio emission peak
reaches  the X-ray emission  decreases by a factor of $10^6$ from
its initial value $\sim 10^{44}$ erg s$^{-1}$.

Thus, we expected a radio flux  with the value $\sim 10^8$ Jy when
the X-ray flux is already unseen. Galactic nuclei with such a
brightness are commonly presented in the Universe \citep{galaxies}
and what is interesting that in some cases there is no activity in other
energy ranges. That falls well into our model though these
speculations cannot be considered as direct evidences in favor of
our model.

DOC and VAD  are partly supported by the RFBR grant 08-02-00170-a,
the NSC-RFBR Joint Research Project ¹ RP09N04 and
09-02-92000-HHC-a and by the grant of a President of the Russian
Federation "Scientific School of Academician V.L.Ginzburg".
KSC is supported by a GRF grant of Hong Kong Government.
CMK is supported by the Taiwan National Science Council
grants NSC 96-2112-M-008-014-MY3 and NSC 98-2923-M-008-001-MY3.
WHI is supported by the Taiwan National Science Council grants
NSC 96-2752-M-008-011-PAE and NSC 96-2111-M-008-010.

We would like to mention that the referee's report was very useful
for us, and we thank him for his comments. DOC and VAD thank the
Institute of Astronomy of NCU for hospitality during their stay
there.

\end{document}